\documentclass[AMA,STIX1COL]{WileyNJD-v2}

\articletype{Research Article}%

\begin{document}

\title{The impact of clustering binary data on relative risk towards a study of inferential methods}

\author[1]{Gopal Nath}
\author[2]{Krishna K. Saha}
\author[3]{Suojin Wang}

\authormark{Nath, Saha, and Wang}

\address[1]{\orgdiv{Department of Mathematics and Statistics}, \orgname{Murray State University, 6C-13 Faculty Hall,  Murray}, \orgaddress{\state{KY 42071}, \country{USA}}}

\address[2]{\orgdiv{Department of Mathematical Sciences}, \orgname{Central Connecticut State University, New Britain}, \orgaddress{\state{CT 06050}, \country{USA}}}

\address[3]{\orgdiv{Department of Statistics}, \orgname{Texas A\&M University, College Station}, \orgaddress{\state{TX 77843}, \country{USA}}}

\corres{Krishna K. Saha, Department of Mathematical
	Sciences, Central Connecticut State University,
	1615 Stanley Street, NewBritain,CT 06050,
	USA. \\ \email{sahakrk@ccsu.edu}}

%\presentaddress{This is sample for present address text this is sample for present address text}

\abstract[Summary]{In epidemiological cohort studies, the relative risk (also known as risk ratio)
	is a major measure of association to summarize the results of two treatments or exposures. Generally, it measures the relative change in disease risk as a result of treatment application. Standard approaches to estimating relative risk available in common software packages may produce biased inference when applied to correlated binary data collected from longitudinal or clustered studies. In recent years, several methods for estimating the risk ratio for correlated binary data have been published, some of which maintain a well-controlled coverage probability but do not maintain an appropriate interval width or the interval location to measure the balance between distal and mesial noncoverage probabilities accurately or, vice versa. This paper develops efficient and straightforward inference procedures for estimating a confidence interval for risk ratio based on a hybrid method. In general, the hybrid method combines two separate confidence intervals for two single risk rates to form a hybrid confidence interval for their ratio. Additionally, we propose the procedures for constructing a confidence interval for risk ratio that directly extends recently recommended methods for correlated binary data by building on the concepts of the design effect and effective sample sizes typically used in representative sample surveys. In order to investigate the performance of these proposed methods, we conduct an extensive simulation study. To demonstrate the utility of our proposed methods, we present three examples from real-life applications, comparing the side effects of low-dose tricyclic antidepressants with a placebo, the efficacy of the treatment group in a teratological experiment, and the efficiency of the active drugs in curing infection for clinical trials. 
}


	\maketitle
	
%	\footnotetext{\textbf{Abbreviations:} ANA, anti-nuclear antibodies; APC, antigen-presenting cells; IRF, interferon regulatory factor}

	%\include{introduction}
	\section{Introduction\label{intro}}
	It is common to encounter correlated binary data in a wide range of biomedical applications. For example, consider the tricyclic antidepressants study that was first introduced by Furukawa et al.\cite{Furukawa_2003}. 
	The tricyclic antidepressant is a first-generation antidepressant that is used to treat depression, obsessive-compulsive disorder, and chronic pain\cite{Chockalingam_2019}. In each group, 16 studies were included, each considered a cluster \cite{Bakbergenuly_2019}. A total of 667 eligible patients were included in the treatment group and 591 eligible patients were included in the placebo group with mean cluster sizes of 42 and 37 respectively. This randomized trial was designed primarily to assess the side effects of low dosage tricyclic antidepressants with placebo in acute depression. Consider another study that Paul conducted in 1982\cite{Paul1982}. In this toxicological study, results relate to litters of different sizes, each of which exhibits a number of abnormalities due to the control group and medium dose. Accordingly, individuals within the same litter respond similarly and are therefore correlated. This study has a moderate number of clusters with an average cluster size of 8. One purpose of such a study is to evaluate whether the treatment has any effect on the incidence of abnormalities in live foetuses. Let $\gamma_i, i = 1, 2,$ be the proportion of abnormalities in live foetuses who received the $i$th treatment. Then the risk ratio,  $\eta = \gamma_1/\gamma_2$, will determine whether the treatment affects the incidence of abnormalities in live foetuses. We wish to develop some efficient inference procedures for estimating $\eta = \gamma_1/\gamma_2$ for such a design. The confidence interval that does not consider the intraclass correlation will generate an inaccurate estimate of coverage probability, expected width and distal noncoverage as a percentage of the total coverage probability, resulting in an inappropriate confidence interval. This problem makes it impossible to apply the standard analysis which is based on the assumption that observations within clusters are independent\cite{Paul2010} \cite{KK_2019}. Several important measures of association can be used to determine this inference problem, including risk difference, odds ratio, and   risk ratio \cite{Lui_2004}.  An inference can be drawn from one measure of association or another, depending on how the study was designed. A risk difference (RD) is defined as the difference between the risk of an event for exposed and unexposed groups. The risk difference can be measured in public health issues in order to determine the magnitude of excess mortality attributed to each disease. A risk ratio is used to measure the strength of the association between a given disease and a suspected risk factor in toxicological and epidemiological  studies. The risk ratio and odds ratio are essentially equal if the proportions of control and treatment groups are less than 0.1. As long as the proportions of control and treatment groups are not too small, the risk ratio is often regarded as a more reliable measure of effectiveness than the odds ratio. Additionally, in applications, the risk ratio can be interpreted directly\cite{Bakbergenuly_2019}. In the case of correlated binary data, a variety of inference procedures have been developed to determine the risk difference and the risk ratio; however, little attention has been given to the extension of the numerous existing procedures. In this paper, we focus on minimizing this gap by extending some of the recommended procedures for a single proportion to the ratio of the proportions in two treatment groups.
	
	Lui presented four methods for calculating a confidence interval for a risk ratio\cite{Lui_2004}. In 2010, Paul and Zaihra\cite{Paul2010} assessed Lui's four methods by comparing them with a method using an estimator of the variance of the ratio estimator\cite{Cochran_1977} as well as with another using a sandwich estimator for the variance of the regression estimator\cite{Zeger_1986} and  concluded that the method (MR3) that focuses on variance of ratio estimator performs best overall for constructing confidence intervals for risk ratios. In general, the method performs reasonably well, but it may not be appropriate for cases with higher risk ratios and large intracluster correlation differences. Moreover, the expected length using MR3 increases substantially when the number of clusters is small and the intraclass correlation is greater than 0.5\cite{Paul2010}. Consider, for instance, the simulated data set presented in Table  \ref{Table:Unexpected_width} which includes both treatment and control groups. In this situation, the expected length of the confidence interval using MR3 is unexpectedly large. Note that the intracluster correlations for the treatment and control groups are 0.35 and 0.12, respectively. In addition, the confidence interval does not exist for  MR3 in some scenarios. According to Zaihra and Paul\cite{Paul2010}, MR3 rejects 16511 samples before it takes 10,000 good samples, based on particular parameter settings. For example, consider the simulated data set for treatment and control groups in Table \ref{Table:unsatisfy_interval}. In this instance, the confidence interval using MR3 does not exist. Real-life examples may contain these types of data, so MR3 would not be applicable to such instances. 
	
	\FloatBarrier
	\begin{center}
		\begin{table*}[ht]
			\caption{Simulated data for the treatment and control groups.  Each group consists of 20 clusters with a cluster size of 100. \label{Table:Unexpected_width}}
			\begin{tabular}{llllllllllllllllllllll}
				Dose Groups & \\
				\hline 
				Treatment &	 $Y_1$& 41  & 0   & 46  & 59  & 0   & 38  & 0   & 0   & 0   & 2   & 0   & 6   & 0   & 5   & 41  & 11  & 18  & 20  & 69  & 65  \\
				
				Control&$Y_2$& 0   & 0   & 8   & 2   & 15  & 0   & 0   & 0   & 0   & 0   & 0   & 20  & 0   & 0   & 0   & 7   & 0   & 0   & 0   & 0  \\
				\hline
			\end{tabular} 
		\end{table*}
	\end{center}
	\FloatBarrier

	\FloatBarrier
	\begin{center}
		\begin{table*}[ht]
			\caption{Simulated data for the treatment and control groups. \label{Table:unsatisfy_interval}}
			\begin{tabular}{llllllllllllllllll}
				Dose Groups & \\
				\hline 
				Treatment &	$n_1$ & 12 & 11 & 10 & 9 & 11 & 10 & 10 & 9  & 9  & 5  & 9 & 7  & 10 & 6 & 10 & 7  \\
				&$Y_1$ & 4  & 0  & 1  & 0 & 1  & 0  & 0  & 0  & 1  & 0  & 0 & 0  & 1  & 1 & 3  & 1  \\
				Control&	$n_2$ & 13 & 12 & 9  & 9 & 8  & 8  & 13 & 12 & 10 & 10 & 9 & 13 & 5  & 7 & 10 & 10 \\
				&$Y_2 $ & 0  & 1  & 0  & 0 & 0  & 0  & 0  & 0  & 0  & 1  & 0 & 0  & 0  & 0 & 0  & 0 \\
				\hline 
			\end{tabular}
		\end{table*}
	\end{center}
	\FloatBarrier
	
	Furthermore, even though MR3 maintains a well-controlled coverage probability in most scenarios, it does not maintain Newcombe's\cite{Newcombe_2011} recommended ratio of distal noncoverage probability to mesial noncoverage probability for the confidence interval. In order to address these issues, we propose several alternative confidence interval procedures for the risk ratio by modifying the recommended binomial intervals for correlated binary data based on the hybrid technique and the effective sample size and its adjusted number of successes. In this paper, we construct several explicit asymptotic two-sided confidence intervals (CIs) for $\eta = \gamma_1/\gamma_2$ using the method of variance of estimates recovery (MOVER) proposed by Zou and Donner\cite{Zou_Donner_2008} and also known as the square-and-add method introduced by Newcombe\cite{Newcombe_1998}. The basic idea is to recover variance estimates required for the proportion ratio from the confidence limits for single proportions. The CI estimators for a single proportion, which are incorporated with the MOVER, will include the CIs proposed by Donner and Klar\cite{Donner_Klar_1993}, Lee and Dubin\cite{Lee_Dublin_1994}, Jung and Ahn (2000)\cite{Jung_Ahn_2000}, and Saha et al.\cite{KK_2016_Cov}.

	Katz et al.,\cite{Katz_1978} in 1978, estimated a confidence interval for the risk ratio, which was later modified by Walter and Pettigrew\cite{Walter_1975}\cite{Pettigrew_1986}. Based on the effective sample size and its adjusted success rate, we extend this confidence interval. In Section \ref{Section:Effective_Sample_Size}, we describe how we adjust the effective sample size and the number of successes using three different variances. Newcombe\cite{Newcombe_2001} proposed a method for calculating confidence intervals for the risk ratio which is based on inverse sine transformation. Koopman\cite{Koopman_1984} proposed an asymptotic scoring method for determining confidence intervals for the risk ratio. Lui\cite{Lui_2004} developed an asymptotic confidence interval for the risk ratio based on Fieller's Theorem. The improvement of the method is obtained by applying the Katz logarithm transformation\cite{Katz_1978}. Additionally, Lui proposed an asymptotic confidence interval based on Bailey's\cite{Lui_2004}. This paper extends these four existing methods, each with three different effective sample sizes and adjusted success rates to construct confidence intervals for the risk ratio, as discussed in Section \ref{Section:Effective_Sample_Size}.
	
	The following section describes these 16 methods for estimating the confidence interval for the risk ratio for the correlated data. MR3 will also be discussed as a baseline method. A comprehensive simulation study is presented in Section \ref{simulation_section} by showing 8 out of the 17 methods, based on the simulation results and the properties of the confidence intervals. Furthermore, we illustrate three examples of real-world applications to validate our proposed models in Section\ref{section:illustrations}.  Finally, some concluding remarks are given in Section \ref{section_conclusion}.

	\section{Confidence interval for the Ratio of Two Success Rates}
	\label{Risk_Ratio_CI}
	Suppose that we independently sample $m_i$ clusters from the $i$th treatment, $i =1, 2$. Let $n_{ij}$ be the number of
	individuals in the $j$th cluster, $j =1, \dots, m_i$, who received the $i$th treatment.
	Furthermore, suppose that $Y_{ij}$ of the $n_{ij}$ individuals are total successes by the $i$th treatment. Under the
	usual assumption $Y_{ij}|p_{ij}$ follows binomial $(n_{ij}, p_{ij})$, where $p_{ij}$ is the probability that an individual in
	the $j$th cluster was cured by the $i$th treatment. We further assume that the binomial probability $p_{ij}$ is a random variable having mean $\gamma_i$ and variance $\gamma_i(1-\gamma_i)\theta_i$. The unconditional
	mean and variance of $Y_{ij}$ are then $n_{ij}\gamma_i$ and $n_{ij}p_{ij}(1-p_{ij})[1+(n_{ij}-1)\theta_i]$, respectively. Note
	that the parameter $\gamma_i$ is the success rate of an individual who received the $i$th treatment and the parameter $\theta_i$ is the common intraclass correlation between the binary observations within each cluster in the $i$th group. In this article, the parameter of interest is $\eta = \gamma_1/\gamma_2$. In particular, we would like to construct explicitly simple but efficient confident interval procedures of $\eta$.
	The following subsections explain how we develop effective and reliable confidence interval procedures for the risk ratio between two proportions for correlated binary data.
	
	\subsection{Hybrid Method \label{sec_hybrid}}
	A hybrid method is called the method of variance estimates recovery (MOVER), which was first developed by Zou and Donner\cite{Zou_Donner_2008}. Newcombe\cite{Newcombe_1998} also referred to it as the square-and-add method. Using this approach, two separate confidence intervals for the two individual success rates are combined to construct a single confidence interval for the ratio of two success rates, $\eta = \gamma_i/\gamma_2$. In order to construct a confidence interval for $\eta$, first consider  a $100(1 - \alpha)\%$ CI for $\gamma_1 - \gamma_2$, where $\gamma_1$ and $\gamma_2$ denote any two parameters of interest. Let $\hat{\gamma}_1$ and $\hat{\gamma}_2$ be two estimates of $\gamma_1$ and $\gamma_2$ from two samples independent of each other, respectively. By the Central Limit Theorem, a $100(1 - \alpha)\%$ CI for $\gamma_1 - \gamma_2$ is given by $(L^*, U^*)$, where
	$$L^* = \hat{\gamma}_1 - \hat{\gamma}_2 - z_{\alpha/2}\sqrt{\mbox{var}(\hat{\gamma}_1) + \mbox{var}(\hat{\gamma}_2)}~~\mbox{and}~~U^* = \hat{\gamma}_1 - \hat{\gamma}_2 + z_{\alpha/2}\sqrt{\mbox{var}(\hat{\gamma}_1) + \mbox{var}(\hat{\gamma}_2)}.$$
	However, this procedure performs well only if the sample size is sufficiently large, or if the sampling distributions of $\hat{\gamma}_i$ $(i = 1, 2)$ are close to a normal distribution. From the equations above, it can be shown that $L^*$ and $U^*$
	can be regarded as the minimum and maximum parameter values that satisfy
	$$\frac{[(\hat{\gamma}_1 - \hat{\gamma}_2) -L^*]^2}{\mbox{var}(\hat{\gamma}_1) + \mbox{var}(\hat{\gamma}_2)} =  z_{\alpha/2}^2~~\mbox{ and }~~\frac{[U^*-(\hat{\gamma}_1 - \hat{\gamma}_2)]^2}{\mbox{var}(\hat{\gamma}_1) + \mbox{var}(\hat{\gamma}_2)} =  z_{\alpha/2}^2,$$
	respectively. Suppose a $100(1 - \alpha)\%$ CI for $\gamma_i$ is $(l_i, u_i), i=1,2$, where $l_i = \hat{\gamma}_i - z_{\alpha/2}\sqrt{\mbox{var}(\hat{\gamma}_i)}$ implies $\widehat{\mbox{var}}(\hat{\gamma}_i) = (\hat{\gamma}_i - l_i)^2/z_{\alpha/2}^2$ under $\gamma_i\approx l_i$. Similarly, $u_i = \hat{\gamma}_i + z_{\alpha/2}\sqrt{\mbox{var}(\hat{\gamma}_i)}$ implies $\widehat{\mbox{var}}(\hat{\gamma}_i) = (u_i - \hat{\gamma}_i)^2/z_{\alpha/2}^2$ under $\gamma_i\approx u_i$. Based on the possible values $(l_1, u_1)$ of $\gamma_1$ and $(l_2, u_2)$ of $\gamma_2$, the values closest to the minimum $L$ and maximum $U$ are $l_2 - u_1$ and $u_2-l_1$, respectively. As a result, to set $L$ with $\gamma_2\approx l_2$ and $\gamma_1\approx u_1$, we have
	$\widehat{\mbox{var}}(\hat{\gamma}_1) + \widehat{\mbox{var}}(\hat{\gamma}_2) =  (u_1 - \hat{\gamma}_1)^2/z_{\alpha/2}^2 + (\hat{\gamma}_2 - l_2)^2/z_{\alpha/2}^2$, which gives
	\begin{eqnarray}
		L^* \approx \hat{\gamma}_1 - \hat{\gamma}_2 - \sqrt{(\hat{\gamma}_2 - l_2)^2 + (u_1 - \hat{\gamma}_1)^2}. \label{eq.1}
	\end{eqnarray}
	Similarly, we have
	\begin{eqnarray}
		U^* \approx \hat{\gamma}_1 - \hat{\gamma}_2 + \sqrt{(u_2 - \hat{\gamma}_2)^2 + (\hat{\gamma}_1 - l_1)^2}.
		\label{eq.2}
	\end{eqnarray}
	Now, let $(L, U)$ be the $(1 - \alpha)100\%$ confidence interval for $\eta = \gamma_i/\gamma_2$, that is,
	$$P(L \leq \gamma_1/\gamma_2 \leq U) = 1 - \alpha.$$
	Equivalently,
	$$P(\gamma_1 - U\gamma_2 \leq 0 \leq \gamma_2 - L\gamma_1) = 1 - \alpha.$$
	For fixed $L$ and $U$, we apply (\ref{eq.1}) to $\gamma_1 - U\gamma_2$ and (\ref{eq.2}) to $\gamma_2 - L\gamma_1$ and by setting $L^* = 0$ and $U^* = 0$, we obtain an approximate $(1 - \alpha)100\%$ confidence interval for $\eta = \gamma_i/\gamma_2$ as
	\begin{eqnarray}
		L = \frac{\hat{\gamma}_1\hat{\gamma}_2 - \sqrt{(\hat{\gamma}_1\hat{\gamma}_2)^2 - u_2(2\hat{\gamma}_1 - l_1)l_1(2\hat{\gamma}_2 - u_2)}}{u_2(2\hat{\gamma}_2- u_2)} \label{eq.3}
	\end{eqnarray}
	and
	\begin{eqnarray}
		U = \frac{\hat{\gamma}_1\hat{\gamma}_2 + \sqrt{(\hat{\gamma}_1\hat{\gamma}_2)^2 - u_1(2\hat{\gamma}_1 - u_1)l_2(2\hat{\gamma}_2 - l_2)}}{l_2(2\hat{\gamma}_2- l_2)}. \label{eq.4}
	\end{eqnarray}
For the equations above, the restrictions $u_2(2\hat{\gamma}_2- u_2) >0 , ~ l_2(2\hat{\gamma}_2- l_2) > 0, ~ (\hat{\gamma}_1\hat{\gamma}_2)^2  > u_2(2\hat{\gamma}_1 - l_1)l_1(2\hat{\gamma}_2 - u_2),  \text{and} ~ (\hat{\gamma}_1\hat{\gamma}_2)^2 >  u_1(2\hat{\gamma}_1 - u_1)l_2(2\hat{\gamma}_2 - l_2)$  are required in order to avoid negative or complex values for $L$ and $U$.

	It is easily seen that to obtain a $100(1 - \alpha)\%$ MOVER based confidence interval for
	$\eta = \gamma_1/\gamma_2$ using Equations (\ref{eq.3}) and (\ref{eq.4}), one needs two separate $100(1 - \alpha)\%$ confidence
	intervals: ($l_1$, $u_1$) for $\gamma_1$ and ($l_2$, $u_2$) for $\gamma_2$. In the study of confidence intervals for clustered binary data, Saha et al.\cite{KK_2016_Cov} examined the problem of confidence intervals for a single proportion. Based on their analysis, they recommended the Wilson score method.

	\subsubsection{The Wilson score interval}
	The natural estimator of $\gamma_i$ $(i=1, 2)$ can easily be obtained as the overall sample proportion $\hat{\gamma}_i = Y_{i.}/n_{i.}$, where $Y_{i.} = \sum_{j=1}^{m_{i}}Y_{ij}$ and $n_{i.} = \sum_{j=1}^{m_i}n_{ij}$. The variance of $\hat{\gamma}_i$ is given by var($\hat{\gamma}_i$) = $\gamma_i(1 - \gamma_i)\xi_i/n_{i.}$, where $\xi_i = \sum n_{ij}[1+(n_{ij}-1)\theta_i]/n_{i.}$. Using the Central Limit Theorem, it can be shown that $n_{i.}^{1/2}(\hat{\gamma}_i - \gamma_i)/\sqrt{\gamma_i(1 - \gamma_i)\hat{\xi}_i}$ converges in distribution to the standard normal distribution as $m = {\rm min} \{m_1, m_2\} \rightarrow \infty$, where $\hat{\xi}_i$ is obtained by replacing $\theta_i$ by its estimate $\hat{\theta}_i$. Then the approximate $100(1-\alpha)$\% Wilson confidence interval for $\gamma_i$ is the root of the quadratic equation
	$$P(n_{i.}(\hat{\gamma}_i - \gamma_i)^2/[\gamma_i(1 - \gamma_i)\hat{\xi}_i] \leq z^2_{\alpha/2}) = 1 - \alpha.$$
	After some straightforward algebra, it can be obtained as
	$$\mbox{WI}:~~(l_i, u_i)=\tilde{\gamma}_i \pm \frac{z_{\alpha/2}}{\tilde{n}_{i.}}\sqrt{n_{i.}\hat{\gamma}_i(1 - \hat{\gamma}_i)\hat{\xi}_i +\frac{\hat{\xi}_i^2z^2_{\alpha/2}}{4}},$$
	where
	$$\tilde{\gamma}_i = \frac{n_{i.}\hat{\gamma}_i + 0.5\hat{\xi}_iz^2_{\alpha/2}}{n_{i.} + \hat{\xi}_iz^2_{\alpha/2}} = \frac{Y_{i.} + 0.5\hat{\xi}_iz^2_{\alpha/2}}{n_{i.} + \hat{\xi}_iz^2_{\alpha/2}}~~\mbox{and}~~ \tilde{n}_{i.} = n_{i.} + \hat{\xi}_iz^2_{\alpha/2}.$$
	It is worthwhile to note here that for non-clustered data when there is no cluster effect, that is, $\theta_i = 0$ (or $\xi_i = 1$), the same intervals are produced (see, for example, Newcombe\cite{Newcombe_1998}). The estimate $\hat{\theta}_i$ can be obtained using the analysis of variance (ANOVA) method, which is given by $\hat{\theta}_{i}^a = (BMS_i - WMS_i)/[BMS_i + (n_i^* - 1)WMS_i]$, where $BMS_i=[\sum_j Y_{ij}^2/n_{ij}-(\sum_j Y_{ij})^2/\sum_jn_{ij}]/(m_i -1)$ and $WMS_i=[\sum_j Y_{ij}-\sum_j Y_{ij}^2/n_{ij}]/\sum_j(n_{ij}-1)$ are the between mean-squared and within mean-squared errors, respectively, and $n_i^* = [(\sum_jn_{ij})^2 - \sum_jn_{ij}^2]/[(m_i -1)\sum_jn_{ij}]$. Therefore, one can obtain Wilson CIs for $\gamma_i$ $(i=1, 2)$ using the above interval $(l_i, u_i)$  by plugging the ANOVA estimate of $\theta_i$ in the equation for $\hat{\xi}_i$ above. We denote this interval as HB$_1$.

	\subsection{Effective sample size based on consistent estimator of variance for correlated binary data \label{Section:Effective_Sample_Size}}
	In general, if $Y_i. \sim B(n_i,\gamma_i) $,  the variance of $\hat \gamma_i$ is $\gamma_i (1-\gamma_i)/n_i$. However, due to intracluster correlation, the variance of the observed response rate inflates the usual variance \cite{KK_2019}.
	We estimate effective sample size, $n_i$, and its adjusted number of success, $Y_i$, based on a complex variance estimator by incorporating unequal and equal weights, and a robust estimator  of the variance of a  ratio estimator. Here is an overview of the three variance estimators and corresponding effective sample sizes: 
	
	i) The complex variance estimator based on equal weights can be estimated as \cite{Jung_Ahn_2000}\cite{Lee_Dublin_1994}\cite{KK_2019}\cite{Rao_Scottt_1992}: 
	$$v^{eq}_i=\hat{v}(\hat{\gamma}_i^\zeta) = \frac{1}{m_i(m_i-1)}\sum_{j=1}^{m_i} (\hat{\gamma}_{ij} - \hat{\gamma}_i^\zeta)^2,$$ 
	where $\hat{\gamma}_i^\zeta = \sum_{j=1}^{m_i}\hat{\gamma}_{ij}/m_i$,
	and the degrees of freedom adjusted effective sample size  as \cite{Korn_Graubard_1998}:  
	\begin{equation}
		n_{i.}^{eq} = \frac{\hat{\gamma}_i^\zeta(1-\hat{\gamma}_i^\zeta)}{v^{eq}_i}\left(\frac{t_{\alpha/2,~ n_{i.}-1}}{t_{\alpha /2, ~ m_{i.}-1 }}\right)^2,
		\label{eq_adj_df_Korn}
	\end{equation}
	where $t_{1-\alpha/2, ~ d}$ is the $1-\alpha/2$ quantile of a $t$ distribution with $d$ degrees of freedom.

	ii) The complex variance estimator based on optimal weights can be estimated as \cite{Jung_Ahn_2000}\cite{Lee_Dublin_1994}\cite{KK_2019}\cite{Rao_Scottt_1992}: 
	$$v^{op}_i=\hat{v}(\hat{\gamma}_i^\xi) = \frac{1}{m_i-1}\sum_{j=1}^{m_i} \xi_{ij}(\hat{\gamma}_{ij} - \hat{\gamma}_i^\xi)^2,$$ 
	where $$ \xi_{ij} =\frac{\left(\frac{n_{ij}}{ 1+(n_{ij}-1)\theta_i}\right)} {\sum_{j=1}^{m_i} \left(\frac{n_{ij}}{ 1+(n_{ij}-1)\theta_i}\right)}~ \geq 0, \text{  } \hat{\gamma}_i^\xi = \sum_{j=1}^{m_i}\hat{\gamma}_{ij} \xi_{ij},$$
	$\sum_{j=1}^{m_i}\xi_{ij} = 1$ and $\theta_i$ is the intraclass correlation. 
	Similarly, using Equation (\ref{eq_adj_df_Korn}), we can estimate the degrees of freedom adjusted effective sample size $n_{i.}^{op}$ based on $v^{op}_i$  and $ \hat{\gamma}_i^\xi $. 
	
	iii) Paul and Zaihra\cite{Paul2008, Paul2010} also developed  the variance of a ratio estimator  to estimate the variance of $\hat\gamma_i$
	$$v^{re}_i= \frac{m_i}{m_i-1}  \sum_{j=1}^{m_i} (Y_{ij}-n_{ij}\hat{\gamma}_i)/ n_i^2.$$
	We can then estimate the effective sample size by  $ n^{re}_{i.}=\hat{ \gamma}_i(1-\hat{\gamma}_i) / v^{re}_i$.

	\subsubsection{Confidence Interval for risk ratio based on modified Katz log and its adjusted sample size}
	According to Katz et al.\cite{Katz_1978}, the ratio of proportions is approximately normally distributed. The confidence interval for the risk ratio, $\eta $, based on the Delta method is given by
	$$\exp \left(  \text{log}(\hat{\eta}) \pm z_{\alpha/2}\sqrt{\frac{1}{Y_{1.}} + \frac{1}{Y_{2.}} + \frac{1}{n_{1.}} + \frac{1}{n_{2.}}} \right)~~~~ Y_{i.}= \sum_{j=0}^{m_i} Y_{ij} \ne 0,  ~ n_{i.}= \sum_{j=0}^{m_i} n_{ij} \ne 0. ~~~~ \text{and }~~ n_{i.}\ne Y_{i.}. $$
	A modified version of Walter and Pettigrew's Katz based confidence interval\cite{Walter_1975}\cite{Pettigrew_1986} is given as:
	
	$$\exp \left( \text{log}(\hat{\eta}^*) \pm z_{\alpha/2}\sqrt{\frac{1}{Y_{1.} + 0.5} + \frac{1}{Y_{2.} + 0.5} + \frac{1}{n_{1.} + 0.5} + \frac{1}{n_{2.} + 0.5}} \right),  ~~ ~ n_{i.} \ne Y_{i.} ,$$
	where $$\eta^* = \frac{(Y_{1.} + 0.5)(n_{2.} + 0.5)}{(Y_{2.} + 0.5)(n_{1.} + 0.5)}.$$
	Moreover, a modified Katz log confidence interval for the risk ratio based on $n_{i.}^{eq}$ and its adjusted number of success $Y_{i.}^{eq}= \gamma_i  n_{i.}^{eq}$ is given as:
	
	$$\exp \left( \text{log}(\hat{\eta}^{eq}) \pm z_{\alpha/2}\sqrt{\frac{1}{Y_{1.}^{eq} + 0.5} + \frac{1}{Y_{2.}^{eq} + 0.5} + \frac{1}{n_{1.}^{eq} + 0.5} + \frac{1}{n_{2.}^{eq} + 0.5}} \right),  ~~ ~ n_{i.}^{eq} \ne Y_{i.}^{eq} $$
	where $$\hat{\eta}^{eq} = \frac{(Y_{1.}^{eq} + 0.5)(n_{2.}^{eq} + 0.5)}{(Y_{2.} ^{eq}+ 0.5)(n_{1.}^{eq} + 0.5)}.$$
	Henceforth, we will refer to this method as MK1. 
	Similarly, we estimate modified Katz log confidence intervals for the risk ratio based on ($n_{i.}^{op}, Y_{i.}^{op}$) and ($n^{re}_{i.}, Y^{re}_{i.}$) and refer to them as MK2 and MK3.

	\subsubsection{Confidence Interval Based on modified Inverse Hyperbolic Sine}
	Based on inverse hyperbolic sine transformation, Newcombe\cite{Newcombe_2001} obtained a $100(1-\alpha$)\% confidence interval for the risk ratio as follows:
	\begin{equation}
		\exp \left(\text{log}(\hat{\eta}) \pm 2 sinh^{-1}\left(\frac{z_{\alpha/2}}{2}\sqrt{\frac{1}{Y_{1.}} + \frac{1}{Y_{2.}} + \frac{1}{n_{1.}} + \frac{1}{n_{2.}}}\right)\right) ~~~ Y_{i.}, n_{i.} \ne 0 ~~ \text{and}~~ n_{i.}\ne Y_{i.} ,
		\label{eq_hyperbolic}
	\end{equation}
where $sinh(x)=(e^x-e^{-x})/2$.

	We estimate three confidence intervals by replacing ($n_{i.}, Y_{i.}$) in Equation (\ref{eq_hyperbolic}) with three sets of effective sample sizes and the adjusted numbers of successes  ($n_{i.}^{eq}, Y_{i.}^{eq}$),  ($n_{i.}^{op}, Y_{i.}^{op}$), and ($n^{re}_{i.}, Y^{re}_{i.}$). We refer to these intervals as IH1, IH2, and IH3.

	\subsubsection{Confidence interval based on modified Koopman asymptotic score}
	For the ratio of proportions, Koopman\cite{Koopman_1984} proposed an asymptotic score confidence interval.  Let  
	\begin{equation}
		\Psi(\eta) = \frac{(Y_{1.} - n_{1.}\Lambda)^2}{n_{1.}\Lambda(1 - \Lambda)}\left\{1 + \frac{Y_{1.}(\eta -\Lambda)}{n_{2.}(1 - \Lambda)}\right\},
		\label{eq_Koopman_1}
	\end{equation}
	where
	$$\Lambda = \frac{\eta(n_{1.} + Y_{2.}) + Y_{1.} + n_{2.} - \sqrt{[\eta(n_{1.} + Y_{2.}) + Y_{1.} + n_{2.}]^2 - 4\eta N(Y_{1.} + Y_{2.})}}{2N}.$$
	An asymmetric confidence interval  for the risk ratio is defined as ($\eta_l$, $\eta_r$), which is solved by a numerical procedure of the following equation:
	\begin{equation}
		\Psi (\eta) = \chi^2_{1,1-\alpha}.
		\label{eq_Koopman_2}
	\end{equation}
	In Equations (\ref{eq_Koopman_1}) and (\ref{eq_Koopman_2}), we replace the value of ($n_{i.}, Y_{i.}$) with three sets of effective sample sizes and the adjusted numbers of successes  ($n_{i.}^{eq}, Y_{i.}^{eq}$),  ($n_{i.}^{op}, Y_{i.}^{op}$), and ($n^{re}_{i.}, Y^{re}_{i.}$)  as discussed above. We refer to these methods as KA1, KA2, and KA3.

	\subsubsection{Confidence interval based on modified Delta and Katz asymptotic method}
	
	Lui \cite{Lui_2004} developed a confidence interval for the risk ratio using Delta \cite{Bishop_1975} method as follows:
	\begin{equation}
		\text{max} \left( \hat{\eta} \pm  z_{\alpha/2} \sqrt{ \widehat{\text{var}}(\hat{\eta})}, 0 \right),
	\end{equation}
	where $ \widehat{\text{var}}(\hat{\eta}) = \hat{\eta}^2 [(1-\hat{\gamma}_1)/(n_1\hat{\gamma}_1)+ (1-\hat{\gamma}_2)/(n_2\hat{\gamma}_2)]$, and  $z_\alpha$ is the upper 100$\alpha$th percentile of the standard normal distribution. However, the estimated risk ratio can be skewed if both the sample sizes $n_i$ and probabilities of success $\gamma_i$ are small. In such a case, the Katz logarithm transformation\cite{Katz_1978} often improves the asymptotic confidence interval:
	\begin{equation}
		\hat{\eta}  \exp\left(\pm ~ z_{\alpha/2} \sqrt{ \widehat{\text{var}}(\log( \hat{\eta}))} \right),
	\end{equation}
	where 
	\begin{equation}
		\widehat{\text{var}}(\log( \hat{\eta})) = (1-\hat{\gamma}_1)/(n_1\hat{\gamma}_1)+ (1-\hat{\gamma}_2)/(n_2\hat{\gamma}_2).
		\label{eq_KD}
	\end{equation}
	In Equation (\ref{eq_KD}), we replace the value of ($n_{i.}, Y_{i.}$) with three sets of effective sample sizes and the adjusted numbers of successes  ($n_{i.}^{eq}, Y_{i.}^{eq}$),  ($n_{i.}^{op}, Y_{i.}^{op}$), and ($n^{re}_{i.}, Y^{re}_{i.}$)  and refer to these methods as DK1, DK2, and DK3, respectively.

	\subsubsection{Confidence interval based on modified Fieller and Bailey asymptotic method}
	\label{method_FB}

	According to Lui\cite{Lui_2004}, an asymptotic confidence interval was developed for the risk ratio using Fieller's Theorem, as follows:
	\begin{equation}\label{eq_Paul_first}
		[\text{max} \left(  (b - \sqrt{b^2-ac})/a, 0 \right), (b + \sqrt{b^2-ac})/a]; ~~~ a>0, ~\text{and} ~ b^2-ac >0,
	\end{equation}
	where 
	\begin{equation} \label{eq_FB_Paul}
		\begin{split}
			a & = \hat{\gamma}_2^2- z^2_{\alpha/2} \hat{\gamma}_2 (1- \hat{\gamma}_2 )/ n_2,  \\
			b& = \hat{\gamma}_1 \hat{\gamma}_2,\\
			c & = \hat{\gamma}_1^2- z^2_{\alpha/2} \hat{\gamma}_1 (1-\hat{\gamma}_1 )/ n_1.  \\
		\end{split}
	\end{equation}
	As a result of Bailey's suggestion, the asymptotic confidence interval for the risk ratio is presented as follows in order to reduce the skewness of the sampling distribution\cite{Lui_2004}:
	\begin{equation}
		[\text{max} \left(  ((B - \sqrt{B^2-AC})/A)^3, 0 \right), ((B + \sqrt{B^2-AC})/A)^3]; ~~~ A>0, ~\text{and} ~ B^2-AC >0,
	\end{equation}
	where 
	\begin{equation} \label{eq_FB}
		\begin{split}
			A & = \hat{\gamma}_2^{2/3}- z^2_{\alpha/2}(1- \hat{\gamma}_2 )/  (9 n_2 \hat{\gamma}_2 ^{1/3} ),  \\
			B& = (\hat{\gamma}_1 \hat{\gamma}_2)^{1/3},\\
			C & = \hat{\gamma}_1^2- z^2_{\alpha/2}(1- \hat{\gamma}_1 )/  (9 n_1 \hat{\gamma}_1 ^{1/3} ).   \\
		\end{split}
	\end{equation}
	In Equation (\ref{eq_FB}), we replace the value of $n_{i.}$ with three  effective sample sizes   $n_{i.}^{eq}, n_{i.}^{op},  \text{and} ~   n^{re}_{i.}  $  and refer to these methods as FB1, FB2, and FB3, respectively.

	Zaihra and Paul\cite{Paul2010} developed this method by modifying Equation (\ref{eq_FB_Paul})  in the following manner:
	\begin{equation} 
		\begin{split}
			a & = \hat{\gamma}_2^2- z^2_{\alpha/2} v^{re}_2, \\
			b& = \hat{\gamma}_1 \hat{\gamma}_2,\\
			c & = \hat{\gamma}_1^2- z^2_{\alpha/2} v^{re}_1.  \\
		\end{split}
	\end{equation}
	Using these $a , b,  \text{and}~ c$  and Equation (\ref{eq_Paul_first}), we can obtain a resulting asymptotic confidence interval.  Zaihra and Paul referred to it as MR3\cite{Paul2010}.

	In the following sections we will evaluate and compare 17 ways of constructing confidence intervals (five approaches with three different combinations of effective sample sizes and its effective success rates). 
	A list of 15 out of the 17 ways is given in Table \ref{acronyms} together with their acronyms for ease of reference. Further, hybrid method HB1, as discussed in section \ref{sec_hybrid} , will be included in the study along with the baseline method MR3. 
	
	\begin{table}
		
		\caption{Abbreviations for 15 confidence interval estimators\label{acronyms}}
		
		\begin{tabularx}{\textwidth}{
				l
				>{\centering}X
				>{\centering}X
				>{\centering\arraybackslash}X
			}
			\toprule
			& \multicolumn{3}{c}{Effective $n$ and its adjusted $Y$ by}\\
			\cmidrule{2-4}
			Method &  $ v^{eq}$ & $ v^{ow}$& $ v^{re}$\\
			\midrule
			
			Modified Katz log  &  MK1 & MK2 & MK3\\ 
			
			Modified Inverse Hyperbolic Sine &  IH1 & IH2 & IH3\\
			Modified Koopman Asymptotic Score &  KA1 & MA2 & KA3	\\
			Modified Delta \& Katz &  DK1 & DK2 & DK3\\
			Modified Fieller and Bailey  &  FB1 &FB2& FB3\\
			\bottomrule 
		\end{tabularx}
	\end{table}

	\section{Simulation Studies }
	\label{simulation_section}
	In this section, we investigate the performance of the small and moderate sample behavior of the proposed methods in terms of expected coverage probability and expected interval width using the pre-assigned confidence level of  95\%. Furthermore, we consider the interval location property as well, to ensure symmetry between the distal and mesial  noncoverage probabilities. In addition to the 16 confidence intervals discussed in Section \ref{Risk_Ratio_CI}, we consider MR3 as a baseline, which was proposed by Paul and Zaihra\cite{Paul2010}  and discussed in Subsection   \ref{method_FB}. We will select parameter combinations so that the generated data sets of two treatment groups will cover all possible scenarios in the three real-world examples.

	\subsection{Parameter selection and data generation}
	
	For simplicity, we consider two treatment groups with the numbers of clusters $n_1=n_2= 20, 30, 50$ and cluster sizes $m_1=m_2= 5, 50, 100$. We used a baseline proportion,  $\gamma_1$ $=0.2$, along with $\eta=1, 1.25 , 1.5, \text{and}~ 2$. The intraclass correlation coefficients between two treatment groups $(\theta_1, \theta_2)$ were set to (0.1, 0.1), (0.1, 0.25), (0.2, 0.2), and (0.2, 0.25). We generated data sets from beta-binomial distributions by combining each set of parameters for both treatments, allowing a total of 10,000 good replications in which the confidence interval existed for all 17 methods discussed above.

	\subsection{Results}
	
\begin{figure}
		\vspace{-8ex}
		\begin{center}
			\includegraphics[height=3.4in,width=7in]{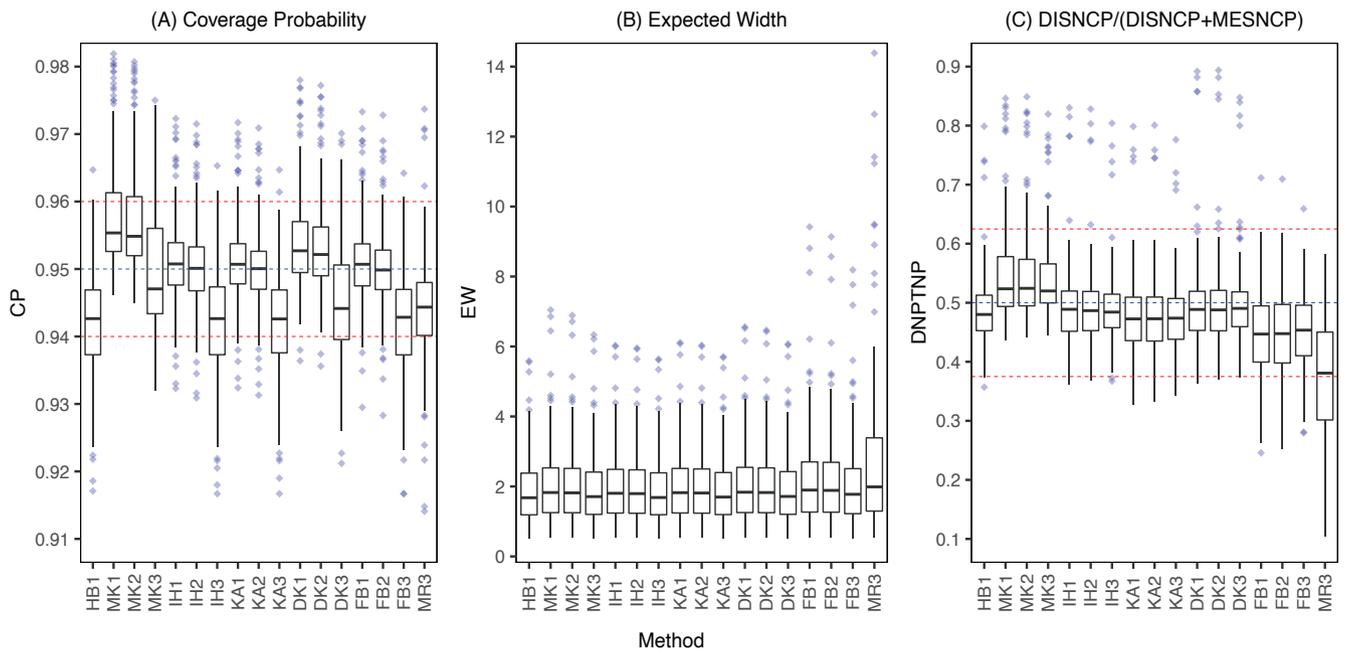}
		\end{center}
		\caption{The estimated coverage probabilities, expected interval widths, and distal noncoverages of the 95\% confidence intervals for $\eta$  of the 17 methods with all parameter combinations.  }
		\label{cov_ew_d}
	\end{figure}

\begin{figure}
		
		\begin{center}
			\includegraphics[height=7in,width=7in]{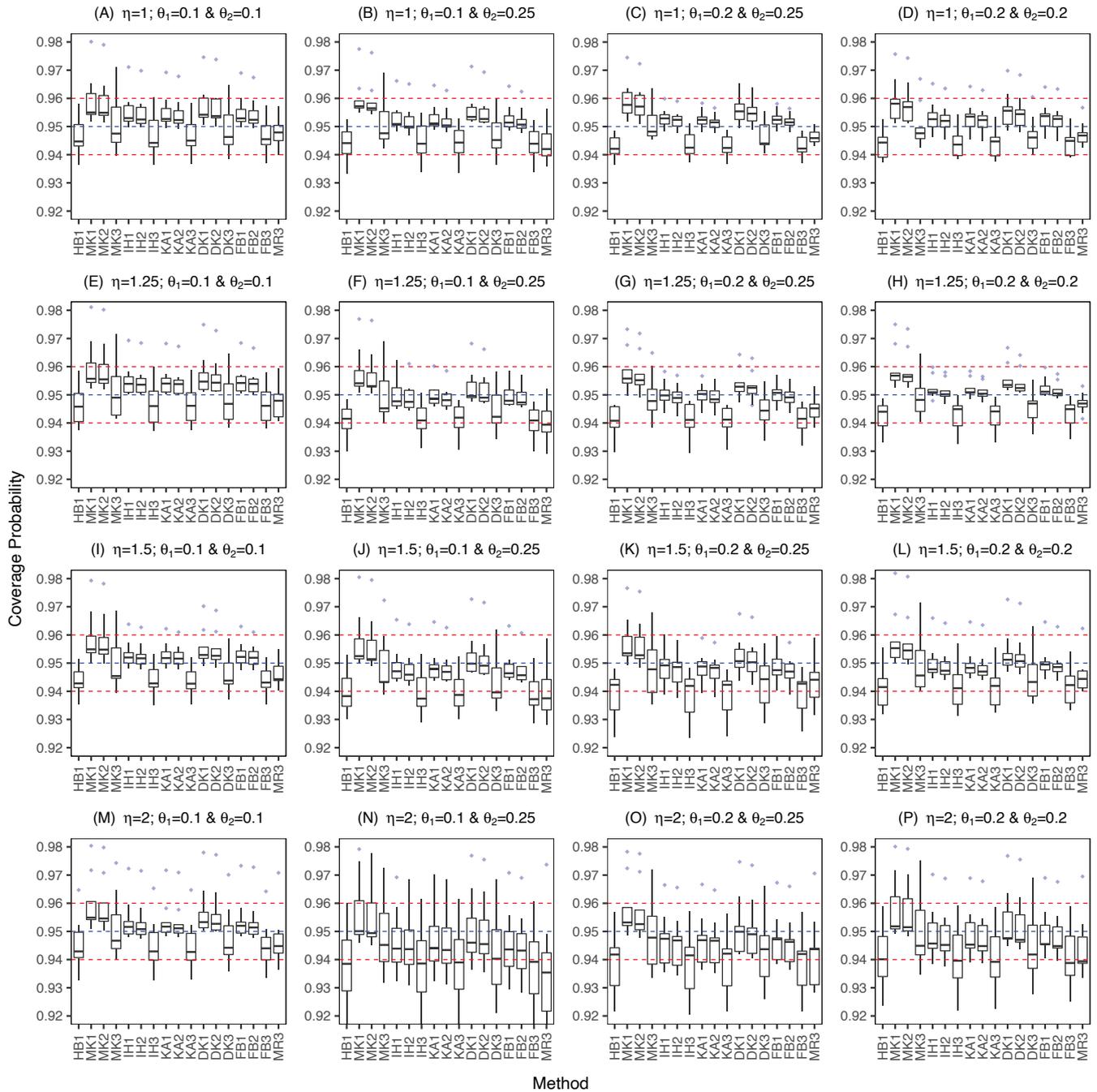}
		\end{center}
		\caption{The estimated coverage probabilities of the 95\% confidence intervals for $\eta$ of the 17 methods. Each plot contains 9 combinations of $n_1=n_2= 20,30, 50$  and $m_1=m_2=5, 50, 100$. In contrast, the 16 plots correspond to different combinations of  $\eta$, $\theta_1$, and $\theta_2$. } 
		\label{cov_RR_PHI}
	\end{figure}

	The observed coverage probability (CP), the expected interval width (EW), and distal noncoverage as a proportion of the total noncoverage probability (DNPTNP) for two-sided confidence intervals ($l$, $u$) for $\eta = \gamma_1/\gamma_2$ were obtained by \cite{KK_2016_Cov} \cite{KK_2013_Mesial} \cite{Newcombe_2011}
	$$\mbox{CP} = \frac{\sum_{t=1}^{10,000}I(l_t\leq  \eta  \leq u_t)}{10,000} ,$$
	where $I = 1$ if $l_t\leq \eta \leq u_t$, and $I$ = 0, otherwise,
	$$~~~ \mbox{EW} = \frac{\sum_{t=1}^{10,000}(u_t - l_t)}{10,000},~~~\text{and}$$
	$$ \mbox{DISNCP} = \frac{\sum_{t=1}^{10,000}J( \eta > u_t)}{10,000}, ~~~~ \mbox{MESNCP} = \frac{\sum_{t=1}^{10,000}K(\eta  <  l_t)}{10,000},$$
	where $J = 1$ if $ \eta > u_t$, and $J$ = 0, otherwise,
	$K = 1$ if $ \eta <  l_t$, and $K$ = 0, otherwise,
	$$\mbox{DNPTNP}= \frac{\mbox{DISNCP}}{\mbox{ DISNCP+MESNCP } }~~~\mbox{and} ~~~ \mbox{EW} = \frac{\sum_{t=1}^{10,000}(u_t - l_t)}{10,000}.$$
	The boxplots in Figure \ref{cov_ew_d}  depict the observed CPs, EWs, and DNPTNPs for the 17 methods for various parameter settings. Each boxplot contains a combination of $ (N_n=3) \times (N_m=3) \times (N_\eta=4) \times (N_{\theta_1:\theta_2}=4) = 144$ parameter. In order to observe how these methods perform overall, we also report the median of the CPs, EWs, and the DNPTNPs in Table \ref{table:cov_median}. According to this table, MK3, IH1, KA1, KA2, DK1, DK2, FB1, FB2, and IH2 have median CPs near 0.95, and MK1, MK2, IH3, KA3, DK3, FB3, HB1, and MR3 have median CPs slightly off from 0.95, but all the median CPs are well controlled. The expected width of all of the methods is similar. As can be seen from the expected width plot, MR3's expected width intervals are remarkably large compared with those of the other EWs. It has also been reported by Zaihra and Paul\cite{Paul2010} in their original paper that when the number of clusters is small and the intraclass correlation is high, the expected width by MR3 is substantially larger. The remaining 16 methods have very similar expected widths. However, FB1 and FB2 have slightly wider expected widths. MK3, IH3, KA3, DK3, FB3, and HB1 have shorter widths than the others, which is indicative of their superior precision. All of the DNPTNPs are very close to 0.5 except MR3. However, the DNPTNP for MR3 is at the very edge of the recommended range between 0.375 and 0.625. As observed in Figure \ref{cov_ew_d}, the DNPTNP can be far outside this recommended interval in some scenarios. We also observe in Figure \ref{cov_ew_d} that in some scenarios IH3, KA3, DK3, FB3, and HB1 have lower CPs than 0.95, whereas MK1, MK2 have higher CPs.  In almost all areas, IH1, IH2, KA1, KA2, DK1, DK2, FB1, FB2, and MR3 maintain well-controlled CPs. We can also see in Figure \ref{cov_ew_d} that MR3  has the largest expected width of the confidence interval. To investigate the performance of the CPs of each method under different scenarios of risk ratio and intracluster correlation, we provide box plots in Figure \ref{cov_RR_PHI}. Each boxplot contains $(N_n=3) \times (N_m=3) =9 ~~$ parameter combinations under the baseline proportion of $\gamma_1=0.2$. In addition, we observe that the methods IH3, KA3, DK3, FB3, HB1, and the baseline method MR3 struggle to maintain the nominal confidence level for higher risk ratios, unequal and larger intracluster correlations. For simplicity, we select MK3, IH2, KA2, DK2, DK3, FB2, HB1, and MR3 as our baseline methods, and for now, we will focus only on these methods for further analysis. 
	
\begin{figure}
		
		\begin{center}
			\includegraphics[height=5.5 in,width=5.5in]{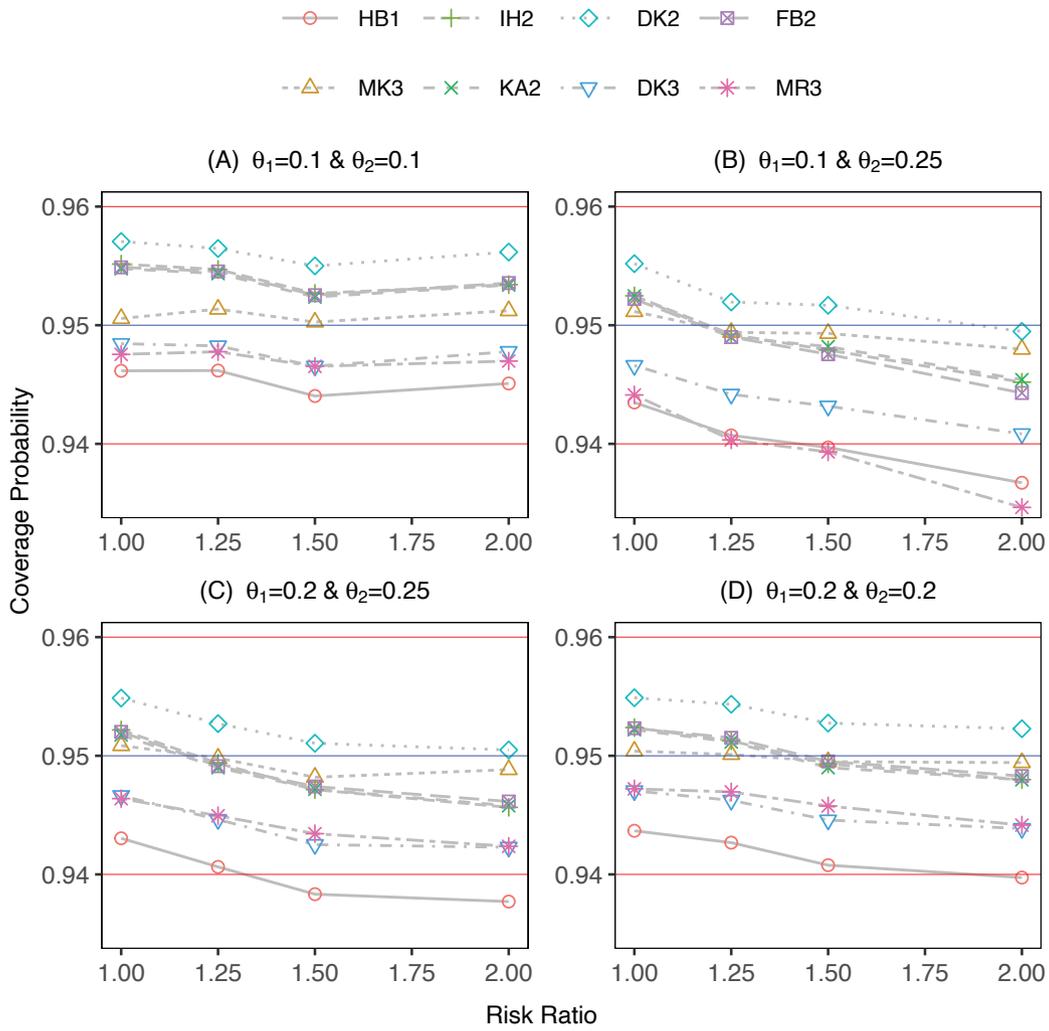}
		\end{center}
		\caption{The estimated coverage probabilities of the 95\% confidence intervals for the selected  8 methods. Each shape contains an average of 9 (combinations of $n_1 = n_2 = n = 20, 30, 50$  and $ m_1 = m_2 = m = 5, 50,  100$) CPs.}
		\label{cov_line_RR_PHI}
		
	\end{figure}

\begin{figure}
		
		\begin{center}
			\includegraphics[height=5.5 in,width=5.5in]{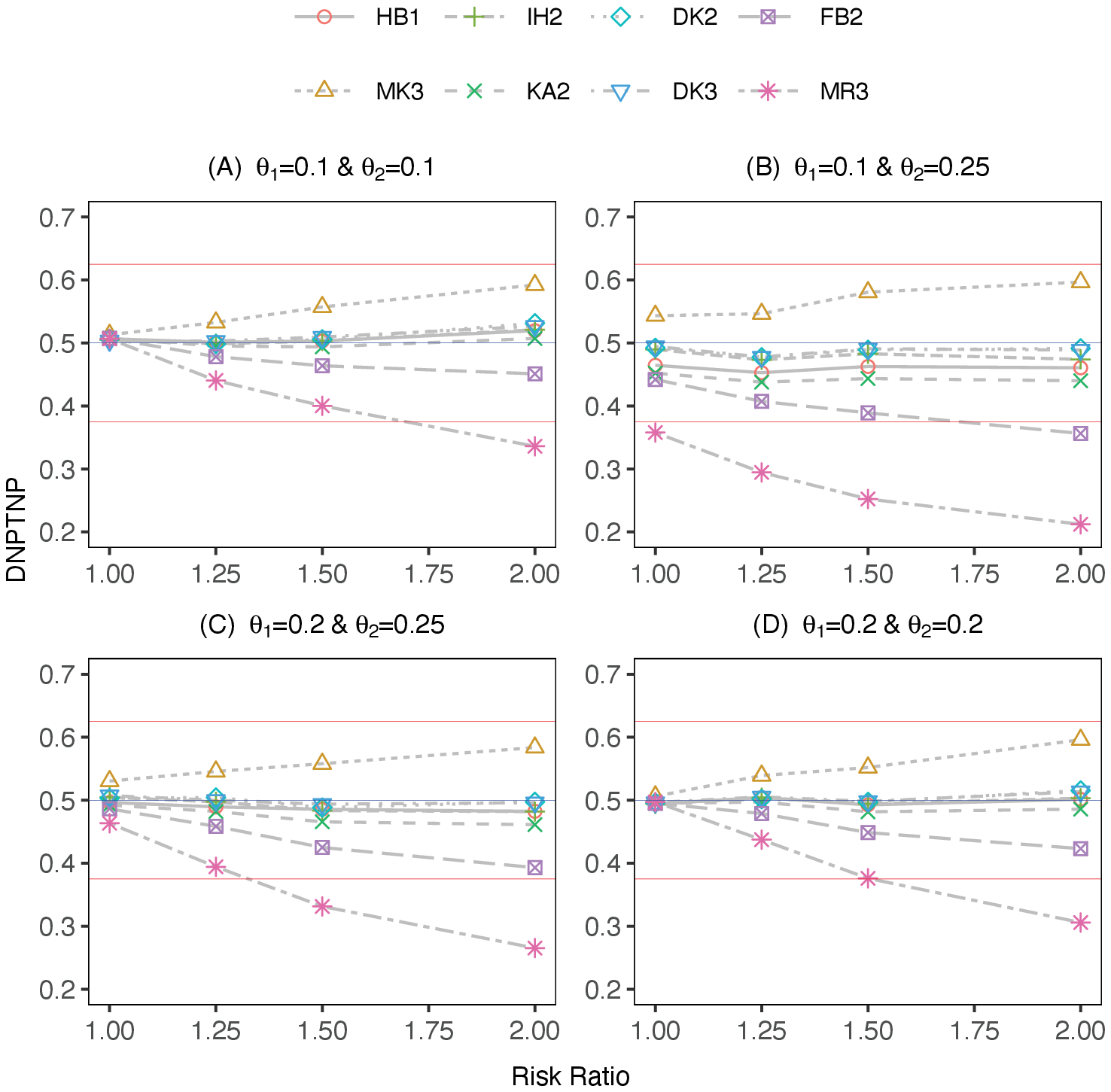}
		\end{center}
		\caption{ Multiple line plots of the distal noncoverages versus the risk ratio for the 95\% confidence intervals for the 8 selected methods. Each shape contains an average of 9 (combinations of $n_1 = n_2 = n = 20, 30, 50$  and $ m_1 = m_2 = m = 5, 50, 100$) DNPTNPs. }
		\label{dn_line_RR_PHI}
	\end{figure}

	The mean CP of each of the four combinations of $\theta_1$ and $\theta_2$ is plotted against the risk ratio in Figure \ref{cov_line_RR_PHI} to observe the association between the  risk ratio  and intracluster correlation. Each shape in a line contains an average of  $(N_n=3) \times (N_m=3) =9 ~~$ CPs. We observe from these plots that all methods except MR3 and HB1 maintain well-controlled coverage probabilities. In the presence of unequal intracluster correlations ($\theta_1=0.1~  \&~  \theta_2= 0.25$), and a risk ratio larger than 1.5, the CPs of MR3 and HB1 are smaller than 0.94. We also observe that HB1 displays liberal CPs for intracluster correlations of 0.2 and 0.25 and a risk ratio more than 1.5. Moreover,  CPs for MK3, DK2, IH2, and Fb2 are very close to the nominal interval level in all of the scenarios. MR3 consistently provides larger EWs in all scenarios. In particular, for $ \theta_1=0.2,  \theta_2=0.25$, and $\eta=2$, MR3 provides an exceedingly large EW value.  To assess the balance between distal (DISNCP) and mesial (MESNCP) noncoverages, we present line plots of DNPTNP=(DISNCP/(DISNCP+MESNCP) in Figure \ref{dn_line_RR_PHI}. Each shape of a line consists of an average of nine DNPTNPs. According to Newcombe\cite{Newcombe_2011}, the interval location is satisfactory if DNPTNP is within a range of 0.375 to 0.625. There are two horizontal red lines in the plot representing these values.  In all scenarios, all of the methods except for MR3 preserve a balance between distal and mesial noncoverages. Moreover, IH2, KA2, DK2, DK3, and HB1 maintain a well-balanced DNPTNP, which is 0.5 in all scenarios.

	\begin{center}
		\begin{table}[t]%
			\centering
			\caption{ Based on 144 parameter combinations, the median cover probabilities (CP), median expected widths (EW), and median of mesial noncoverages  are given for the 17 methods. 	\label{table:cov_median}}
			\begin{tabular*}{500pt}{@{\extracolsep\fill}lccc@{\extracolsep\fill}}
				\toprule
				\textbf{Method} & \textbf{Median CP}  & \textbf{Median EW}  & \textbf{ Median DNPTNP }  \\
				\midrule
				MK1 & 0.955 & 1.826 & 0.524 \\
				MK2 & 0.955 & 1.816 & 0.520 \\
				MK3 & 0.947 & 1.709 & 0.520 \\
				IH1 & 0.951 & 1.804 & 0.489 \\
				IH2 & 0.95  & 1.795 & 0.487 \\
				IH3 & 0.943 & 1.684 & 0.484 \\
				KA1 & 0.951 & 1.821 & 0.473\\
				KA2 & 0.95  & 1.812 & 0.473 \\
				KA3 & 0.943 & 1.696 & 0.474 \\
				DK1 & 0.953 & 1.835 & 0.489 \\
				DK2 & 0.952 & 1.825 & 0.488 \\
				DK3 & 0.944 & 1.713 & 0.490 \\
				FB1 & 0.951 & 1.896 & 0.447 \\
				FB2 & 0.95  & 1.887 & 0.448 \\
				FB3 & 0.943 & 1.775 & 0.454 \\
				HB1 & 0.943 & 1.677 & 0.480  \\
				MR3 & 0.944 & 1.988 & 0.381 \\
				\bottomrule
			\end{tabular*}
			
		\end{table}
	\end{center}

	\begin{figure}
		
		\begin{center}
			\includegraphics[height=5.5 in,width=5.5in]{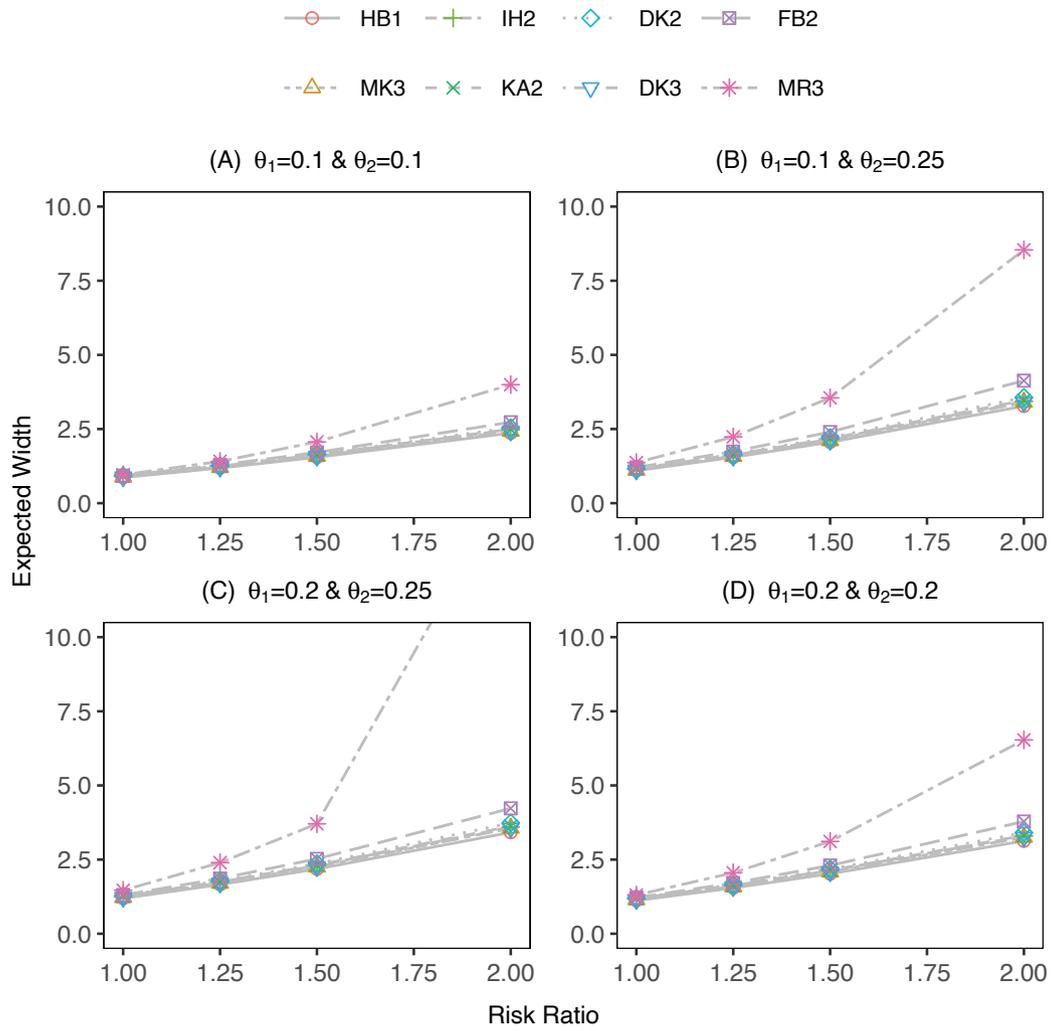}
		\end{center}
		\caption{The expected widths of the 95\% confidence intervals for the selected  8 methods. Each shape contains an average of 9 (combinations of $n_1 = n_2 = n = 20, 30, 50$  and $ m_1 = m_2 = m = 5, 50, 100$) EWs.  }
		\label{ew_line_RR_PHI}
	\end{figure}

	\section{Applications to real-world data}
	\label{section:illustrations}
	In this section, we discuss three real-world examples from biomedical studies.  Our first example is the tricyclic antidepressants study.
	
	%%%%%%%%%%%%%%%%%%%%%%%%%%%%%%%%%%%%%%%%%%%%%%%%%%%%%%%%%%%%%%%
	%%%%%%%%%%%%%%%%%                       Example 1                  %%%%%%%%%%%%%%%%%%%%%%%%%%%
	%%%%%%%%%%%%%%%%%%%%%%%%%%%%%%%%%%%%%%%%%%%%%%%%%%%%%%%%%%%%%%%
	
	\subsection{Example 1: Tricyclic Antidepressants Study}
\begin{figure}[ht]
		\begin{center}
			\includegraphics[height=2.2in,width=4.5in]{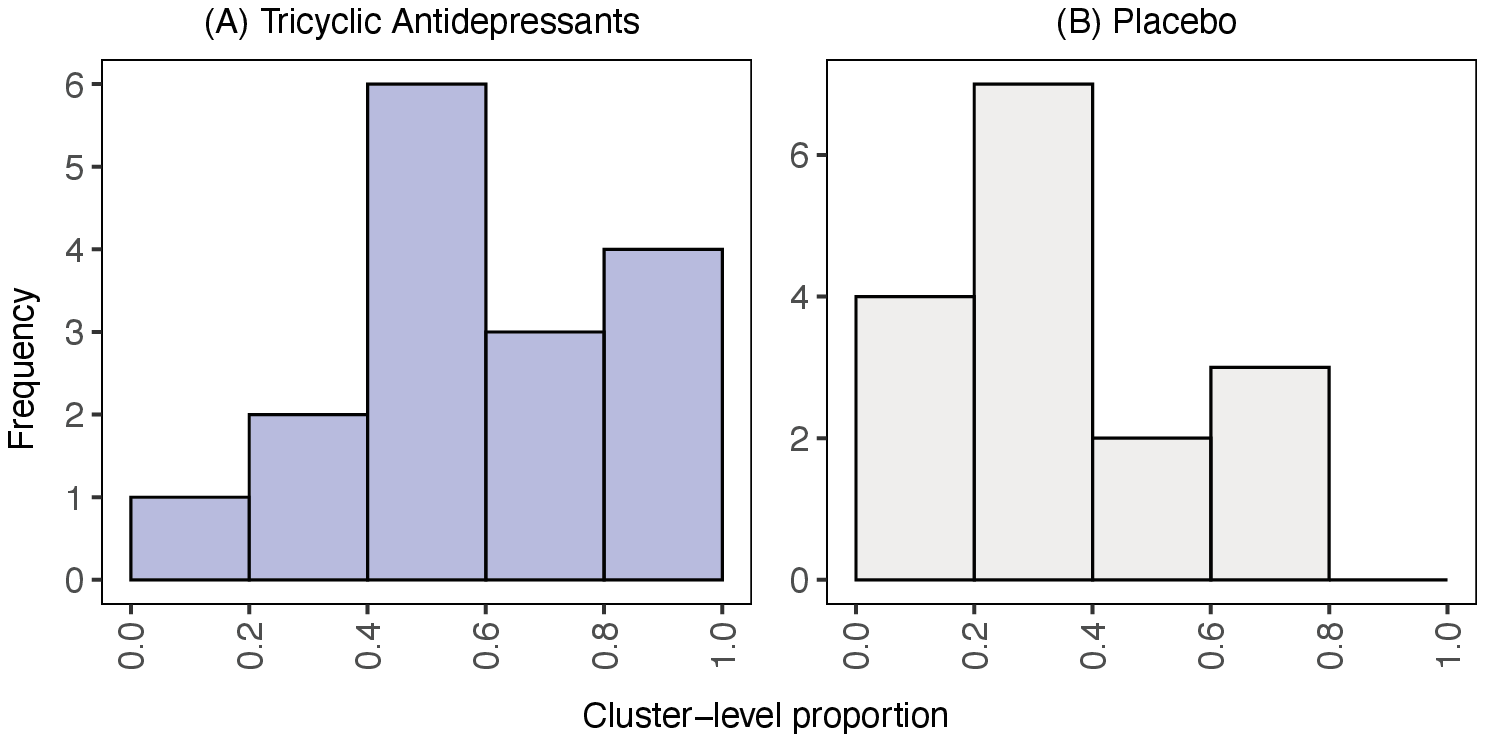}
		\end{center}
		\caption{The distributions of cluster-level proportions for both treatment groups in the tricyclic antidepressant  study}
		\label{prop_dist_ex_antidepressants}
	\end{figure}

	Consider the tricyclic antidepressant study first introduced by Furukawa et al.\cite{Furukawa_2003}, and was subsequently analyzed by Bakbergenuly et al.\cite{Bakbergenuly_2019}. The use of tricyclic antidepressants continues to be widespread around the world. A common criticism of practicing physicians and psychiatrists is that they administer tricyclic antidepressants at too low a dosage to those with depression. In adults with major depressive disorder, tricyclic antidepressants have been associated with a wide range of adverse effects. However, no systematic evaluation of the serious or non-serious adverse effects associated with all types of tricyclic antidepressants has been conducted\cite{Jorgensen_2021}. The study aims to compare the side effects of low-dose tricyclic antidepressants with a placebo. The empirical distributions of cluster level proportions for treatment and placebo are shown  in Figure \ref{prop_dist_ex_antidepressants}. Estimated success probabilities for low dosage tricyclic antidepressants and placebo are 0.604 and 0.390, respectively.  The ANOVA estimates for the intercluster correlation coefficients for the intervention and control groups are 0.169 and 0.164, respectively.  According to Bakbergenuly et al. \cite{Bakbergenuly_2019}, there was a statistically significant difference between the treatment group and the placebo group.  Before we use our proposed methods, we need to confirm whether they are appropriate for this example. By using the same parameter combinations as in this example, we conducted simulations using the technique described in Section \ref{simulation_section}.  We obtained coverage probabilities,  expected widths, and distal noncoverages of the 95\% confidence intervals for the risk ratio by all 8 selected methods based on $\hat{\gamma}_t=0.604$, $\widehat{\eta}={\hat{\gamma}_t}/{\hat{\gamma}_p}= 1.545$ , $n_t=16, m_t=41.68$ ,  $n_p=16$,  and  $m_p=36.93$  as shown in Table \ref{table:prop_dist_ex_antidepressants}.  HB1 and MR3 have liberal CPs, while the remainder have well-controlled CPs(close to 0.95). In our simulation study, we observed similar coverage probabilities for all the methods in Figure \ref{cov_RR_PHI}(L). The expected widths of all the methods are also similar. Each DNPTNP is very close to 0.5, with the exception of MR3, which is 0.406, but within the recommended range between 0.375 and 0.625\cite{Newcombe_2011}. This result supports the use of these methods in this situation, leading to the usage of the confidence intervals of all 8 methods. According to these results, each of the 8 CIs is greater than 1, which suggests that tricyclic antidepressants have a significant effect. All the confidence widths are similar, however as expected from our simulation study, DK3, MK3, and IH2 have shorter widths. By contrast, HB1 followed by MR3 have a larger width. Based on the data analysis in this scenario, it appears that MK3, IH2, DK2, and DK3 are most likely to provide precise and accurate interval estimation.
	\FloatBarrier
	
	\begin{center}
		\begin{table*}[ht]%
			\caption{Test of appropriateness and the 95\% confidence intervals for the risk ratio of the  tricyclic antidepressant and placebo groups using the 8 selected out of the 17 methods.\label{table:prop_dist_ex_antidepressants}}
			\centering
			\begin{tabular*}{500pt}{@{\extracolsep\fill}lcccccc@{\extracolsep\fill}}
				\toprule  
				&\multicolumn{3}{@{}c@{}}{\textbf{Test of appropriateness}} &  \multicolumn{3}{@{}c@{}}{\textbf{The 95\% confidence interval for  $\eta$}} \\\cmidrule{2-4}\cmidrule{5-7}
				\textbf{Method} & \textbf{Coverage Probability}  & \textbf{Expected width}  &   \textbf{DNPTNP} &{\textbf{Lower limit (L)}}  & \textbf{Upper limit (U)}  &  \textbf{Width} \\
				\midrule
				
				HB1 & $\textbf{0.939}^\dagger $& 1.017 & 0.504 & 1.074 & 2.244 & 1.169 \\
				MK3 & 0.947 & 1.027 & 0.553 & 1.084 & 2.176 & 1.092 \\
				IH2 & 0.958 & 1.102 & 0.519 & 1.089 & 2.195 & 1.106 \\
				KA2 & 0.956 & 1.121 & 0.472 & 1.096 & 2.223 & 1.127 \\
				DK2 & 0.958 & 1.108 & 0.519 & 1.087 & 2.199 & 1.112 \\
				DK3 & 0.943 & 1.024 & 0.522 & 1.095 & 2.183 & 1.088 \\
				FB2 & 0.957 & 1.122 & 0.488 & 1.095 & 2.22  & 1.125 \\
				
				MR3 & $\textbf{0.938}^\dagger $ & 1.080  & 0.406 & 1.112 & 2.263 & 1.151 \\

				\bottomrule
			\end{tabular*}
			
			\begin{tablenotes}
				
				\item[$\dagger$] A liberal or conservative CP (less than 0.94 or greater than 0.96 \cite{KK_2019})
			\end{tablenotes}
		\end{table*}
	\end{center}
	
	\FloatBarrier

	%%%%%%%%%%%%%%%%%%%%%%%%%%%%%%%%%%%%%%%%%%%%%%%%%%%%%%%%%%%%%%%
	%%%%%%%%%%%%%%%%%                        Example 2                                   %%%%%%%%%%%%%%%%%%%%%
	%%%%%%%%%%%%%%%%%%%%%%%%%%%%%%%%%%%%%%%%%%%%%%%%%%%%%%%%%%%%%%%

	\subsection{Example 2: Teratological  Experimental  Study}
	We revisit an example of a teratological experiment that was originally conducted by Paul \cite{Paul1982} and subsequently analysed by Saha and Wang\cite{KK_2019}. In this study, the data refer to litters of varying sizes, each litter having a number of abnormalities due to a control group and medium dose. Within the same litter, individuals respond in a similar manner, so they are correlated. 
	
\begin{figure}[ht]
		
		\begin{center}
			\includegraphics[height=2.2in,width=4.5in]{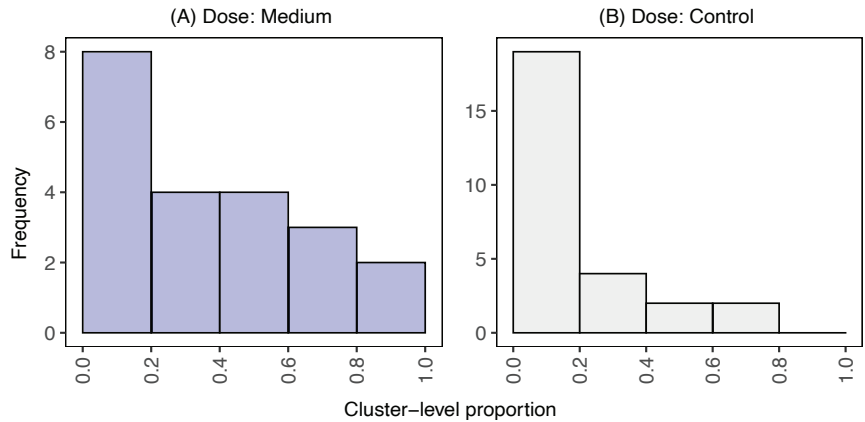}
		\end{center}
		\caption{The distributions of cluster-level proportions for both treatment groups in the teratological  experimental  study}
		\label{prop_dist_ex_tera}
	\end{figure}

	The empirical distributions of cluster level proportions for both treatment groups are shown in Figure \ref{prop_dist_ex_tera} which indicates that the distributions are highly skewed. Estimated success probabilities for medium and control  doses are 0.344 and  0.134, respectively.  ANOVA estimates for the intracluster correlation coefficients for the intervention and control groups are 0.277 and 0.218, respectively.  According to Saha and Wang\cite{KK_2019}, there was a statistically significant difference between the treatment group and the control group. Therefore, we chose this example to evaluate the performance of our proposed method in such a case.  Before we use our proposed methods, we need to confirm whether they are appropriate for this example. By using the same parameter combinations as in this example, we conducted simulations using the technique described in Section \ref{simulation_section}.  We obtained  coverage probabilities,  expected widths, and mesial noncoverages of the 95\% confidence interval  for the risk ratio by all 8 methods based on parameter combination of  $\hat{\gamma}_m=0.344$, $\widehat{\eta}={\hat{\gamma}_m}/{\hat{\gamma}_c}= 2.55$, $n_m=21, m_m=7.19$,  $n_c=27$,  and  $m_c=7.96$  as shown in Table \ref{table:prop_dist_ex_tera}. According to the results, all 8 selected methods have well-controlled coverage probabilities. In accordance with Newcombe\cite{Newcombe_2011}, all the methods except MR3 maintain the location property as their MNPTNPs for the 95\% confidence intervals for $\eta$ are between 0.375 and 0.625  while MR3's MNPTNP is 0.275. Based on the simulation results in Figure \ref{dn_line_RR_PHI}(B), it is evident  that MR3 struggles to maintain the location property for higher risk ratios and for unequal intracluster correlations ($\theta=0.1~ \&~\theta=0.25$). These results support the use of our proposed methods in this situation. However, the use of MR3 in this situation should be regarded critically. None of the 95\% confidence intervals based on the 8 selected methods includes 1, which suggests a significant difference between the medium dose and the control dose (which is consistent with Saha and Wang's \cite{KK_2019} conclusions). As expected, MK3 has the shortest interval width, and MR3 has the largest interval width in this scenario.  On the basis of our evaluation of the appropriateness of the models and the width of the confidence intervals, we conclude that models MK3, IH2, KA2, DK2, DK3, and HB1 provide the most reliable confidence intervals for $\eta$.

	\FloatBarrier
	
	\begin{center}
		\begin{table*}[t]%
			\caption{Test of appropriateness and the 95\% confidence intervals for the risk ratio of the medium  and control dose using the 8 selected out of 17 methods in a teratological experimental study.\label{table:prop_dist_ex_tera}}
			\centering
			\begin{tabular*}{500pt}{@{\extracolsep\fill}lcccccc@{\extracolsep\fill}}
				\toprule  
				&\multicolumn{3}{@{}c@{}}{\textbf{Test of appropriateness}} &  \multicolumn{3}{@{}c@{}}{\textbf{The 95\% confidence interval for  $\eta$}} \\\cmidrule{2-4}\cmidrule{5-7}
				\textbf{Method} & \textbf{Coverage Probability}  & \textbf{Expected width}  &   \textbf{DNPTNP} &{\textbf{Lower limit (L)}}  & \textbf{Upper limit (U)}  &  \textbf{Width} \\
				\midrule
				HB1 & 0.941 & 3.867 & 0.526 & 1.319 & 4.89  & 3.571 \\
				MK3 & 0.952 & 3.99  & 0.587 & 1.336 & 4.696 & 3.36  \\
				IH2 & 0.950 & 4.1   & 0.504 & 1.301 & 5.01  & 3.709 \\
				KA2 & 0.951 & 4.122 & 0.503 & 1.303 & 5.035 & 3.732 \\
				DK2 & 0.955 & 4.201 & 0.516 & 1.285 & 5.074 & 3.789 \\
				DK3 & 0.947 & 4.016 & 0.507 & 1.373 & 4.749 & 3.376 \\
				FB2 & 0.951 & 4.437 & 0.434 & 1.312 & 5.268 & 3.956 \\
				
				MR3 & 0.941 & 5.981 & $\textbf{0.275}^\ddagger$  & 1.413 & 5.558 & 4.145 \\
				
				\bottomrule
			\end{tabular*}
			
			\begin{tablenotes}
				\item[$\ddagger$] DNPTNP lies outside the recommended range between 0.375 and 0.625\cite{Newcombe_2011}
			\end{tablenotes}
		\end{table*}
	\end{center}

	\FloatBarrier

	%%%%%%%%%%%%%%%%%%%%%%%%%%%%%%%%%%%%%%%%%%%%%%%%%%%%%%%%%%%%%%%
	%%%%%%%%%%%%%%%%%                                       Example 3                    %%%%%%%%%%%%%%%%%%%%%
	%%%%%%%%%%%%%%%%%%%%%%%%%%%%%%%%%%%%%%%%%%%%%%%%%%%%%%%%%%%%%%%

	\subsection{Example 3: Active drug's effectiveness in curing infection}
\begin{figure}[ht]
		\begin{center}
			\includegraphics[height=2.2in,width=4.5in]{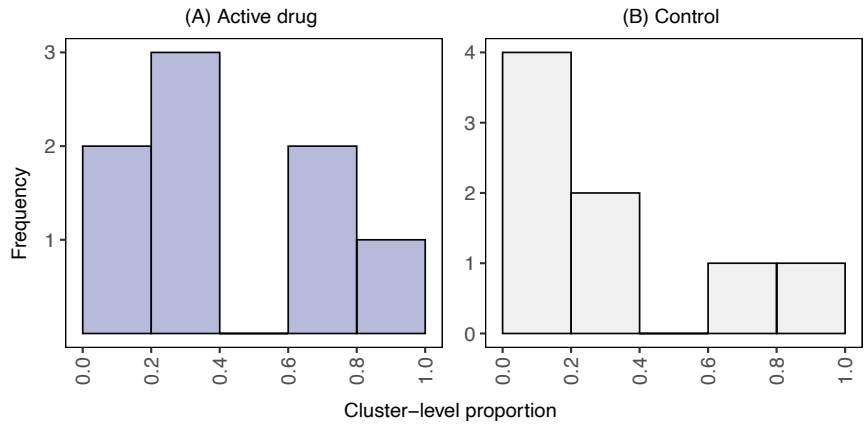}
		\end{center}
		\caption{The distribution   of $\gamma_a$ and $\gamma_c$ for the infection treatment clinical trial.  }
		\label{prop_dist_active_drug}
	\end{figure}

	Consider the example of the multicenter randomized clinical trial, which was first introduced by Beitler and Landis \cite{Beitler1985}. In this clinical trial, an active anti-infective drug was compared with a control drug to determine if it effectively treats infections. In this example, eight clinics are considered clusters. The mean cluster sizes for active and control drugs are 16.25 and 17.87, respectively. The empirical distributions of $\hat \gamma_a$ and $\hat \gamma_c$ are provided in Figure \ref{prop_dist_active_drug}. The ANOVA estimates of the intracluster correlation coefficients for the treatment and control groups are 0.258 and 0.328, respectively. Prior to using these methods to estimate the confidence interval for $\eta$ of this example, a simulation was performed with similar parameter configurations.  Based upon the notation in Section \ref{simulation_section},  we considered  the parameter combination $\hat{\gamma}_a=0.42$, $\widehat{\eta}=\frac{\hat{\gamma}_a}{\hat{\gamma}_c}=1.28$ , $n_a=n_c=8, m_a=16.25$ and $ m_c=17.87$. Under this combination, we obtained the coverage probabilities, expected widths and mesial noncoverages for $\eta$ at the 95\% level.   According to the results in Table \ref{table:active_drug}, MK3  and DK3  have well-controlled coverage probabilities. A liberal coverage probability is observed for HB1 and MR3. IH2, KA2, and FB2 have slightly higher coverage probabilities than the well-controlled CP(0.94, 0.96), while DK3 has a conservative probability. The expected widths of all the methods are similar except for MR3, which is unusually large compared to the other methods. HB1, MK3, IH2, DK2, and DK3 have the DNPTNP values near 0.5 while MR3 has the DNPTNP value within the recommended range although very close to the lower range. Based on these results, we recommend that our proposed methods be used in this scenario. In this situation, however, employing MR3 should be viewed critically. In this example, all the confidence intervals include 1, which suggests the chances of being cured with an active versus control drug are not significantly different. DK3 (2.24) has the shortest interval width followed by  MK3 (2.28),  HB1 (2.395), and IH2 (2.574). In contrast, MR3 (3.14) has the largest width followed by FB2 (2.86).
	We conclude that based on our evaluation of the appropriateness of the models and the widths of the confidence intervals, models MK3, DK3, and IH2 most likely provide reliable confidence intervals for $\eta$.  
	
	\FloatBarrier

	\begin{center}
		\begin{table*}[ht]%
			\caption{Test of appropriateness and the 95\% confidence intervals for the risk ratio of the active and control group using the 8 selected out of  17 methods in a randomized multicenter clinical trial.\label{table:active_drug}}
			\centering
			\begin{tabular*}{500pt}{@{\extracolsep\fill}lcccccc@{\extracolsep\fill}}
				\toprule  
				&\multicolumn{3}{@{}c@{}}{\textbf{Test of appropriateness}} &  \multicolumn{3}{@{}c@{}}{\textbf{The 95\% confidence interval for  $\eta$}} \\\cmidrule{2-4}\cmidrule{5-7}
				\textbf{Method} & \textbf{Coverage Probability}  & \textbf{Expected width}  &   \textbf{DNPTNP} &{\textbf{Lower limit (L)}}  & \textbf{Upper limit (U)}  &  \textbf{Width} \\
				\midrule
				HB1 & $\textbf{0.929}^\dagger$ & 2.439 & 0.482 & 0.586 & 2.981 & 2.395 \\
				
				MK3 & 0.951 & 2.632 & 0.540  & 0.558 & 2.84  & 2.282 \\
				IH2 & $\textbf{0.963}^\dagger$ & 2.995 & 0.473 & 0.533 & 3.107 & 2.574 \\
				KA2 & $\textbf{0.962}^\dagger$ & 3.124 & 0.438 & 0.540  & 3.249 & 2.709 \\
				DK2 & $\textbf{0.968}^\dagger$ & 3.155 & 0.488 & 0.518 & 3.198 & 2.680  \\
				DK3 & 0.940  & 2.587 & 0.501 & 0.585 & 2.830  & 2.245 \\
				FB2 & $\textbf{0.963}^\dagger$ & 3.579 & 0.430 & 0.529 & 3.393 & 2.864 \\
				
				MR3 & $\textbf{0.932}^\dagger$ &$\textbf{11.617}^\S$ & 0.393 & 0.581 & 3.721 & 3.140 \\

				\bottomrule
			\end{tabular*}

			\begin{tablenotes}
				\item[$\dagger$] A liberal or conservative CP (less than 0.94 or greater than 0.96 \cite{KK_2019})
				\item[$\S$] Unexpectedly large width in comparison to the example interval width
			\end{tablenotes}
			
		\end{table*}
	\end{center}
	\FloatBarrier

	\section{Discussion and Conclusions}
	\label{section_conclusion}
	
	This paper proposed 16 methods to construct the confidence intervals for the success ratio $\eta = \gamma_1/\gamma_2$ for a correlated binary data based on
	the hybrid procedure using the two separate CIs for a single proportion and the adjusted effective sample size and its number of successes. The 16 methods presented in this study were compared with the existing method MR3 recommended by Zaihra and Paul\cite{Paul2010}\cite{Paul2008} for estimating confidence intervals for risk ratio for correlated binary data. In order to determine the most appropriate method for constructing confidence intervals for correlated binary data, three properties  were considered: (i) coverage probability, (ii) expected width, and (iii) the ratio of distal noncoverage probability to mesial noncoverage probability. According to the findings of our comprehensive simulation study, for a moderate number of clusters with intraclass correlations ranging from 0.1 to 0.25, the MK3, IH2, KA2, DK2, DK3, FB2, and MR3 maintain a well-controlled coverage probability in almost all scenarios. In addition, we observe that HB1, MR3, and DK3 have a lower coverage probability than the nominal level, particularly for higher risk ratios and higher intracluster correlations. 
	According to our analysis, all the methods except FB2 and MR3 had similar expected widths. Furthermore, when the risk ratio increases, the width of MR3 is noticeably larger than the width of the others. In addition, all the methods except MR3 maintain the range between 0.375 and 0.625, which is recommended by Newcombe\cite{Newcombe_2011}. There are noteworthy findings that indicate the DNPTNPs of the MR3 decrease as the risk ratio increases. Based on our observations, the findings of the examples and the results of the simulation study are similar. The performance of MK3, DK3, IH2, and KA2 was well across all properties for a confidence interval. However, MR3 does not meet the performance expectations in at least one scenario, which means at least one of the three properties of the confidence interval was violated. 
	Based on the simulation and case study results, the proposed MK3, IH2, and DK3 methods generally perform well because, in almost all scenarios, the observed CPs are close to the nominal coverage level, confidence widths are smaller and are congruent with expected widths, and DNPTNPs are close to 0.5. Therefore, we recommend the MK3, IH2, and DK3 procedures for the risk ratio  $\eta$ of correlated binary data from epidemiological cohort studies or other fields.

	\begin{center}
		{\bf Acknowledgments}
	\end{center}

This work was partially supported by a CSU-AAUP University research grant and by the Simons Foundation Mathematics and Physical Sciences - Grant for Collaboration among Mathematicians \#499650. 

\baselineskip=22.5pt

%\nocite{*}% Show all bib entries - both cited and uncited; comment this line to view only cited bib entries;
\bibliography{wileyNJD-AMA}%

\clearpage

\end{document}